\documentclass[final,1p]{elsarticle}
\usepackage{amsmath,amssymb}
\usepackage{amsfonts}
\usepackage[dvipsnames]{xcolor}
\usepackage{graphicx}
\usepackage{caption}
\usepackage{subcaption}
\usepackage{dcolumn}

\usepackage{epstopdf}
\usepackage{color}
\usepackage[utf8]{inputenc}
\usepackage{amsthm}
\usepackage{fontenc}
\usepackage{soul}

\usepackage{bm}
\RequirePackage{slashed}\usepackage{cancel}
\usepackage[normalem]{ulem}

\def\be{\begin{eqnarray} &&} 
\def\nonu{\nonumber \\ &&} 
\def\ee{\end{eqnarray}} 
\usepackage{array,booktabs}
\newcolumntype{C}{>{$\displaystyle}c<{$}}

\def\hef{${\rm ^4 He} $}
\def\het{${\rm ^3 He} $}
\def\tri{${\rm ^3 H} $}
\journal{Physics Letters B}
\begin{document}
\begin{frontmatter}

\title{The  EMC effect for few-nucleon bound systems  in Light-Front Hamiltonian Dynamics}
\author[1]{Filippo Fornetti}
\author[2]{Emanuele Pace}
\author[1]{Matteo Rinaldi}
\ead{matteo.rinaldi@pg.infn.it}
\author[3]{Giovanni Salm\`e}
\author[1]{Sergio Scopetta \footnote{S. Scopetta is presently Science Counsellor at the Italian Embassy in Spain, calle Lagasca 98, 28006 Madrid, Spain. email: sergio.scopetta@esteri.it}}
\author[4]{Michele Viviani}
\address[1]{ Dipartimento di Fisica e Geologia,
Universit\`a degli Studi di Perugia 
and Istituto Nazionale di Fisica Nucleare,
Sezione di Perugia, via A. Pascoli, I - 06123 Perugia, Italy}
\address[2]{Universit\`a di Roma ``Tor Vergata'', 
Via della Ricerca Scientifica 1, 00133 Rome, Italy}
\address[3]{Istituto  Nazionale di Fisica Nucleare, Sezione di Roma, Piazzale A. Moro 2,
00185 Rome, Italy}
\address[4]{Istituto Nazionale di Fisica Nucleare, Sezione di Pisa, Lungarno Pacinotti 43, 56127 Pisa}
\cortext[cor2]{Corresponding author}
\date{\today}

\begin{abstract}
The light-front formalism for a covariant description of the European Muon Collaboration (EMC) effect, already applied to \het, is formally extended to any nucleus, and used for actually calculating the 
$^3$H and \hef~cases. 
The realistic and accurate nuclear description of  few-nucleon
bound systems,  obtained with both phenomenological and chiral potentials, has  been properly combined with the Poincar\'e covariance and  macroscopic locality, automatically satisfying both number of particles and momentum sum rule. While  retaining   the on-mass-shell nucleon structure functions, one is then able to predict a sizable  EMC effect for \hef, as already observed for \het. Moreover, the impact on the EMC effect of both i) the short-range correlations, such as those generated by modern nuclear interactions, and ii) the ratio between the neutron and proton structure functions has been studied. The short-range correlations generated by retaining only the  standard 
nuclear degrees of freedom act on the depth of the minimum in the EMC ratio, while the uncertainties linked to the ratio of neutron to proton structure functions are found to be 
{very} small. These light-front results  facilitates ascribing deviations from experimental data due to genuine QCD effects, not included in a standard nuclear description, and initiating unbiased investigations. 
\end{abstract}
\end{frontmatter}

\section{Introduction}

 After forty years of studies, the European Muon Collaboration (EMC) effect \cite{EuropeanMuon:1983wih}, i.e. the depletion of the  ratio, $R^A_{EMC}(x)$,  between the electron-nucleus and  the electron-deuterium deep inelastic scattering (DIS) cross sections, in the valence region of the Bjorken variable $x$,
is still to be fully understood.
Any attempt to explain the DIS electromagnetic (em) response of the nucleus as the incoherent response of the {\em constituent} nucleons, assumed to be the only active degrees of freedom (dofs), failed to succeed.
A natural question has therefore arisen on how and to what extent the nucleus can be used as a QCD laboratory, with quark and gluon dofs  playing a relevant role.
A 
{complete} answer is still lacking, although progresses have been made, 
 e.g. by evaluating the role of short-range correlations (SRCs) between the nucleons when the nucleus is probed by  high-virtuality photons  \cite{Hen:2016kwk,CLAS:2019vsb,Wang:2020uhj}, and also  by addressing  medium modifications of the nucleons, given the large overlap between their wave functions (see, e.g. Ref. \cite{CLAS:2019vsb,Kulagin:2004ie, Geesaman:1995yd,Cloet:2019mql,Segarra:2019gbp}).
{Hopefully, a wealth of complementary information should stem from}  the study of 
semi-inclusive
and exclusive
processes \cite{Dupre:2015jha},
 planned for light nuclei
in a new generation of experiments
at existing 
\cite{Armstrong:2017zqr}
and future \cite{AbdulKhalek:2021gbh} facilities, like
 the  Electron-Ion Collider (EIC), where measurements with (polarized) $^{\rm 3}$He beams is under development.  For now, recent $^{\rm 3}$He and $^{\rm 3}$H DIS data have been taken at JLab by the MARATHON Collaboration \cite{MARATHON:2021vqu} 
allowing a fresh extraction of the neutron to proton structure functions ratio,
{(SFs)}
{$r(x) = F_2^n(x) / F_2^p(x)$, 
needed to obtain the flavor decomposition of parton distribution
(see also Ref. \cite{Cocuzza:2021rfn})}. 
This experimental scenario has motivated our efforts to quantitatively analyse  the EMC effect.

 In Refs. \cite{Pace:2022qoj,Pace:2020ned}, we have recently proposed and applied to \het~ a new approach for evaluating the EMC ratio, $R^A_{EMC}(x)$. The main features are i)  a 
  description of the interacting nuclear system that takes into account Poincar\'e covariance and  macroscopic locality {(see  \cite{KP})} and ii) an accurate description of \het~ wave function, obtained from a realistic nuclear interaction. The chosen relativistic Hamiltonian dynamics is the light-front  (LF) one, proposed by Dirac in his 1949 seminal paper \cite{Dirac:1949cp}, being the suitable framework for studying the hadron dynamics on the light-cone (such as 
  {the one experimentally investigated in the Bjorken limit). In this way, 
  {the correct support of the longitudinal-momentum distribution and} both baryon number and momentum sum rules are automatically implemented 
  {(see Ref. \cite{Oelfke:1990uy}
  for a LF approach based on an Ansatz for the nuclear part)},
  without introducing ad hoc  normalization constants, and a ratio less than 1 is obtained in the typical interval $0.2<x<0.75$.  Our aim is to embed as many general principles as possible and adopt the most refined description of the nuclear dynamics, still keeping the  on-mass-shell nucleon structure functions. As a result, one can fully describe the well-understood nuclear part and leave for further investigations the analysis of QCD effects to be unambiguously ascribed to partonic dofs. In summary, the rationale of our approach is to provide a baseline that considers only dofs adopted for decades to successfully describe the standard nuclear physics, so that one can address the study of genuine QCD effects, like nucleon swelling or exotic quark configurations, without biases generated by the lack of the Poincar\'e covariance and macroscopic locality. In other words, we re-propose  the conventional description of the EMC effect, but with a degree of sophistication of the nuclear description appropriate to modern times.
  
 The present work is mainly devoted to i) extending the  LF formalism of our approach to any nucleus, and ii) providing a novel calculation of the $^3$H and  \hef~  EMC effects. 
 The $\alpha$ particle is a strongly bound system, and therefore represents a challenging test-bed.  A second aim is the comparison of the EMC ratio for \het, \tri~ and \hef~  calculated by adopting different modern NN and NNN interactions. In particular, we have used  i) the NN charge-dependent Argonne V18  \cite{Wiringa:1994wb} plus the NNN Urbana IX \cite{Pudliner:1995wk} (shortly AV18+3N) and ii) two versions of the Norfolk $\chi$EFT interactions, NVIa+3N and NVIb+3N, that generate three-nucleon forces within the same framework \cite{Piarulli:2016vel,Piarulli:2017dwd,Piarulli:2022ulk} (see also Refs. \cite{Epelbaum:2008ga,Machleidt:2011zz} for a general introduction to $\chi$EFT interactions in nuclear physics). Such a comparison sheds light on the dynamical effect of the {SRCs}, generated by retaining only the standard nuclear dofs, in the EMC region $0.2<x<0.75$.  In particular,  as  has long been known (see, e.g.,  the  studies of the y-scaling in the widest range of $A$ in Ref. \cite{CiofidegliAtti:1990rw}), the tail of the nucleon momentum distributions is strongly affected by the SRCs, and in principle they could  manifest  their relevance also at $x<1$.  Differently from other studies, both experimental and theoretical, which address SRCs in the region $x>1$, (see, e.g., Ref. \cite{Arrington:2022sov} and references quoted there in), here we are interested to see to which extent the EMC ratio is sensitive to the effect of the nuclear interaction, leaving  room  to an  investigation of QCD effects, where partonic dofs  not included in a standard nuclear description  
 play the main role,  to be presented elsewhere.
Finally we have explicitly checked that the ratio of structure functions $F^n_2(x)/F^p_2(x)$ we have extracted in Ref. \cite{Pace:2022qoj} from the MARATHON Coll. data \cite{MARATHON:2021vqu} and the one recently obtained in Ref. \cite{Valenty:2022gqm} by using statistical bootstrap, lead to basically equivalent results for the \hef~EMC effect. This outcome could suggest that the on-mass-shell neutron structure function is reliably related to the proton one, while the latter   is shown to be affected by large uncertainties in the region $x>0.7$ (see Sect. 18 in the  recent  Particle Data Group \cite{Workman:2022ynf}).

In addition to  the LF formalism,  we  exploit the Bakamjian-Thomas (BT) construction of the Poincar\'e generators \cite{Bakamjian:1953kh} for  implementing the Poincar\'e-covariant framework able to embed the successful phenomenology of nuclear physics. Actually, the BT  
 construction 
allows one to make the nonrelativistic
description of nuclei fully compliant with a Poincar\'e-covariant one, since  in this way the Poincar\'e generators obey the standard commutations rules once the mass operator fulfills well-defined constraints.  It turns out that   the nonrelativistic mass operators satisfy {these} conditions, given their dependence on scalar products of the independent momenta at disposal (see Ref. \cite{KP}). 

The main ingredient of our approach is the LF spectral function \cite{DelDotto:2016vkh},  $P^N(\tilde{\kappa},\epsilon)$, i.e. the probability distribution of finding a nucleon with LF momentum $\tilde{\kappa}$ in the intrinsic reference frame of the cluster $[1,(A - 1)]$ and the fully interacting $(A - 1)$-nucleon system with intrinsic energy $\epsilon$. In particular, 
{by using} nonsymmetric intrinsic variables, 
the internal motion of the interacting $(A - 1)$-nucleon system 
{can be disentangled} from that of the struck nucleon and hence 
{one is} able to  implement the macroscopic locality. 
{Notably,}  in the Bjorken limit  the nuclear structure function can be obtained more directly from the nucleus wave function in momentum space, rather than through the cumbersome LF spectral-function \cite{Pace:2022qoj}. 

 In the next section of this Letter, we present the formalism adopted. Then, we present the numerical results and the conclusions in the following two sections. 

\section{The unpolarized structure function $F^A_2(x)$  for  a nucleus}

As shown in  Ref.  \cite{Pace:2022qoj},  in our Poincar\'e covariant framework and 
in  the Bjorken limit, the spin-independent structure function for the $A$-nucleon bound system is 
given by  the following convolution 
\be
F^A_2(x) =  
  ~\sum_N ~    \int_{\xi^B_{min}}^{1} ~ d\xi   ~ 
  F^N_2 \left ({mx \over \xi M_A} \right)  ~ {{f}}^{N}_1(\xi) 
  ~ \quad ,
 \label{eq:F2A}
\ee
where 
$x=Q^2/2m \nu$ is the Bjorken variable, with $Q^2=-q^2$ and $q^\mu\equiv \{\nu,{\bf q}\}$ the virtual photon 4-momentum, $\xi^B_{min}=x m/M_A$, $F^N_2 \left ({mx / \xi M_A} \right)$ is the nucleon structure function, $M_A$ the mass of the nucleus and  $f^{N}_1(\xi) $  the light-cone momentum distribution, that reads 
 \be
f^{N}_1(\xi) = 
 \int d{\bf k}_{\perp} ~ n^N(\xi, {\bf k}_{\perp})
 \label{eq:f1xi}
\ee
with $ n^{N}(\xi,{\bf k}_{\perp})$ the LF {\it spin-independent} nucleon momentum distribution,
 obtained from the LF spectral function, $P^N(\tilde{\kappa},\epsilon)$, after integrating over the energy $\epsilon$ {and}
averaging on the spin direction.  {It can be} written 
as follows 
\be 
n^{N}(\xi,{\bf k}_{\perp})={1 \over 2\pi} \int \prod_{i=2}^{A-1} \left [ d{\bf k}_{i} \right ]~\left |{\partial k_z\over \partial \xi}\right|
~{\cal N}^N({\bf k},{\bf k}_2, \dots~ {\bf k}_{A-1})\,. 
\label{eq:momdis}\ee
Here above, $ {\cal N}^N({\bf k},{\bf k}_2, \dots~ {\bf k}_{A-1})$  is the squared  nuclear wave function in momentum space, with the set of independent Cartesian {3-momenta} describing the relative motion between the $i$-th nucleon and the $A-i$ spectator system, i.e. ${\bf k}$ is the relative 3-momentum with respect to the $A-1$ cluster, ${\bf k}_2$ is relative to the $A-2$ one, down to ${\bf k}_{A-1}$ that is the relative 3-momentum of two nucleons. 
Notice that in a Galilean-invariant framework, these momenta reduce to  the familiar Jacobi momenta.
  
Within the LF framework, recalling the properties of the LF boosts (see, e.g. Ref. \cite{KP}), one introduces the following intrinsic, nonsymmetric transverse momenta and LF-longitudinal ones (let us recall that the LF momenta\footnote{We recall that a generic  LF 4-vector $b$ is defined as: $b=(b^-, \tilde b)$, where $\tilde b=(b^+,  {\bf b}_\perp )$ with $b^\pm = b_0 \pm b^3$ and ${\bf b}_\perp=(b_1,b_2)$.} are conserved) for a system composed by on-shell, interacting $A$ nucleons (see also Ref. \cite{DelDotto:2016vkh})
\be
\hspace{-1cm}{\bf k}_\perp= {\bf p}_{1\perp} -\xi {\bf P}_{\perp}~,
\quad \quad ~~~~~\xi={p^+_1 \over P^+}={k^+\over M_0[1,2,3,\dots ,A]}
\nonu
\hspace{-1cm}{\bf k}_{2\perp}= {\bf p}_{2\perp} -\eta_2{\bf P}_{2;\perp} ~,
\quad \quad \eta_2={p^+_2 \over P^+_2}={k^+_2\over M_0[2,3,\dots ,A]}
\nonu
\hspace{-1cm}{\bf k}_{3\perp}= {\bf p}_{3\perp} -\eta_3 {\bf P}_{3;\perp}~,
\quad \quad \eta_3={p^+_3 \over  P^+_3}={k^+_3\over M_0[3,\dots ,A]}
\nonu \vdots \nonu \vdots \nonu 
\hspace{-1cm}{\bf k}_{(A-1)\perp}= {\bf p}_{(A-1)\perp} -\eta_{(A-1)}{\bf P}_{(A-1);\perp}
~,
~  \eta_{(A-1)}={p^+_{(A-1)} \over  P^+_{(A-1)}}={k^+_{(A-1)}\over M_0[A-1,A]}
\label{eq:general}\ee
where ${\bf P}_\perp=\sum_{i=1}^A {\bf p}_{i\perp}$, ${\bf P}_{i;\perp}=\sum_{j=i}^A {\bf p}_{j\perp}$, $P^+= \sum_{i=1}^A p^+_i$,  $P^+_i= \sum_{j=i}^A p^+_j$, and $M^2_0[i,i+1,i+2,\dots~,A]$, for $(A-1)\ge i\ge 1$, is the free-mass of the $(A-i+1)$-nucleon system
(see below for {its definition}). 
The  nonsymmetric variables disentangle the motion of a particle from the motion of the rest of the system.
These variables were introduced  in Ref. \cite{DelDotto:2016vkh}  where a fully interacting spectator system was considered to obtain the nuclear spectral function
and allow one to implement the macroscopic locality {(see also Ref. \cite{KP})}.

The free mass of $A$ nucleons  is given  by 
\be
M^2_0[1,2,3,\dots~,A]= P^+ ~\sum_{i=1}^A {m^2+|{\bf p}_{i\perp}|^2 \over p^+_i}-|{\bf P}_\perp|^2
\ee
and it can be  
decomposed in a contribution from a nucleon and a $(A-1)$ cluster, which depends upon $(A-2)$ relative 3-momenta, viz.
\be
M^2_0[1,2,3,\dots~,A]={m^2+|{\bf k}_\perp|^2\over \xi}+
{M^2_0[2,3,\dots~,A]+|{\bf k}_\perp|^2\over (1-\xi)} \quad ,
\label{eq:masscl1}
\ee
 once the LF-momenta  in Eq. \eqref{eq:general} are adopted. Plainly, this decomposition holds for {the free-mass of any cluster}, that can be further decomposed in the contribution from  a nucleon and a sub-cluster, etc.
 In general the cluster free-mass can be written for  $(A-1)\ge i\ge 2$ as follows
\be
\hspace{-1.6cm}M^2_0[i,i+1,i+2,\dots,A]={m^2+|{\bf k}_{i\perp}|^2\over \eta_i}+{M^2_0[i+1,i+2,i+3,\dots,A]+|{\bf k}_{i\perp}|^2\over (1-\eta_i)}
\label{deco}
\ee
 
 To proceed, we adopt the frame where $ {\bf P}_{\perp}=0$, and $P^+=M_A$.   
Instead of the plus components, $k^+$ and $ k^+_i$, given in Eq. \eqref{eq:general}, we introduce  a new set of variables to match the Cartesian expression for each 3-momentum. As is well-known \cite{KP}, for on-mass-shell particles one can write  
\be
k_z={1\over 2}~\left(k^+-k^-\right)={1\over 2}~\left(k^+- {m^2+|{\bf k}_\perp|^2\over k^+}\right),
\nonu
k_{iz}={1\over 2}~\left(k^+_i-k^-_i\right)={1\over 2}~\left(k_i^+ - {m^2+|{\bf k}_{i\perp}|^2\over k_{ i}^+}\right)~.
\label{eq:kzdef}\ee
Then, from Eq. \eqref{eq:kzdef}, we get  
\be
E({\bf k})=\sqrt{m^2+|{\bf k}|^2}={1\over 2}
~(k^+ + k^-)~,
\ee 
with the Cartesian 3-momentum ${\bf k}\equiv \{k_z,{\bf k}_\perp\}$, and analogously for the $i$-th LF-momentum.
With the above variables 
one gets another expression of the free mass $M_0[1,2,3,\dots~,A]$ decomposed in a contribution from a nucleon and a $(A-1)$ cluster, viz.
\be
M_0[1,2,3,\dots~,A]=\sqrt{m^2+|{\bf k}|^2}+\sqrt{M^2_0[2,3,\dots~,A]+|{\bf k}|^2}~,
\label{eq:masscl}
\ee
where the  $(A-1)$ cluster has to be considered as an on-mass-shell system, with total mass $M_0[2,3,\dots~,A]$ and recoiling with 3-momentum $(-{\bf k})$. Let us recall that the $\pm$ components of the cluster momentum are
\be
\hspace{-1.6cm}K^+_{2,3,\dots,A}= (1-\xi)~M_0[1,2,3,\dots~,A]~, \quad  K^-_{2,3,\dots,A}= {M^2_0[2,3,\dots~,A]+|{\bf k}_\perp|^2\over K^+_{2,3,\dots,A}}.
\label{eq:cluster}
\ee
In general the cluster free-mass can be written for $(A-1)\ge i\ge 2$ as follows
\be
\hspace{-1.6cm}M_0[i,i+1,i+2,\dots,A]=\sqrt{m^2+|{\bf k}_{i}|^2}+\sqrt{M^2_0[i+1,i+2,i+3,\dots,A]+|{\bf k}_{i}|^2}
\label{deco2}\ee
with ${\bf k}_i\equiv \{k_{iz},{\bf k}_{i\perp}\}$.
 
 For $A=2,~3$ one  has, respectively, 
\be 
M^2_0[1,2]={m^2 +|{\bf k}_\perp|^2\over \xi (1-\xi)}=4 
E^2({\bf k})~,
\ee
\be
M^2_0[1,2,3]={m^2+|{\bf k}_\perp|^2\over \xi}+ {M^2_0[2,3]+|{\bf k}_\perp|^2\over (1-\xi)}=\left(E({\bf k})+E_{2,3}({\bf k})\right)^2~,
\ee
with 
\be
M^2_0[2,3]={m^2 +|{\bf k}_{2\perp}|^2\over \eta_2 (1-\eta_2)   }=4 E^2({\bf k}_2)~,
\ee
and 
$E_{2,3}({\bf k})=\sqrt{M^2_0[2,3]+|{\bf k}|^2}$.
For $A=4$ one has
\be
M^2_0[1,2,3,4]={m^2+|{\bf k}_\perp|^2\over \xi}+ {M^2_0[2,3,4]+|{\bf k}_\perp|^2\over (1-\xi)}=\left(E({\bf k})+ E_{2,3,4}({\bf k})\right)^2~,
\ee
with \be 
M^2_0[2,3,4]={m^2+|{\bf k}_{2\perp}|^2\over \eta_2}+ {M^2_0[3,4]+|{\bf k}_{2\perp}|^2\over (1-\eta_2)}=\left(E({\bf k}_2)+ E_{3,4}({\bf k}_2)\right)^2
\nonu
M^2_0[3,4]={m^2 +|{\bf k}_{3\perp}|^2\over \eta_3 (1-\eta_3)   }=4 E^2({\bf k}_3)~,
\ee and $E_{2,3,4}({\bf k})=\sqrt{M^2_0[2,3,4]+|{\bf k}|^2}$.

Eventually, in Eq. \eqref{eq:momdis}, the Jacobian $\Bigl|\partial k_z/\partial \xi\Bigr|$ can be 
 determined from the definitions in Eqs. \eqref{eq:kzdef}, {\eqref{eq:general} and \eqref{eq:masscl1}}, i.e. 
\be
{\partial k_z\over \partial \xi}={\partial k_z\over \partial k^+}{\partial k^+\over \partial \xi}=
{k^+ + k^-\over 2 k^+}  ~\left(M_0[1,2,\dots,A]+{\xi\over 2 M_0[1,2,\dots,A]}~{\partial M^2_0[1,2,\dots,A]\over \partial \xi} \right) 
\nonu
= {E({\bf k})\over k^+}~{E_{2,3\dots A}({\bf k},{\bf k}_{2},\dots,{\bf k}_{A-1})\over (1-\xi)}~,
\ee
with $E_{2,3\dots A}({\bf k},{\bf k}_{2},\dots,{\bf k}_{A-1})~= \sqrt{M^2_0[2,3,\dots,A]+|{\bf k}|^2}$

In summary (cf. Refs. \cite{DelDotto:2016vkh} and \cite{Alessandro:2021cbg} for the three-body case) one writes the explicit expression of the LF spin-independent nucleon momentum distribution for an $A$-nucleon bound system as follows
\be
n^{\tau}(\xi,{\bf k}_{\perp}) = 
{1 \over 2\pi} ~\prod_{i=2}^{A-1}
 \int d{\bf k}_i ~
{E_{2,3,\dots,A}({\bf k},{\bf k}_{2},\dots,{\bf k}_{A-1})~E({\bf k})\over  k^+ ~ (1- \xi)} 
 \nonu 
\times  ~ {1 \over 2J_A+1}
\sum_{{\cal M}=-J_A}^{J_A}
\sum_\sigma\sum_{\{\sigma_\ell\}}
\sum_{\{\tau_\ell\}}\Bigl | \langle \sigma,\{\sigma_\ell\}; 
\tau,\{\tau_\ell\}; {\bf k},
\{{\bf k}_i\}|A;J_A,\pi_A,{\cal M}, T^A_z
\rangle \Bigr|^2
\label{eq:momdis2} 
\ee
where  $\{\sigma_\ell \}\equiv \sigma_2,\sigma_3,\dots,\sigma_A$ are canonical spins and the same notation is adopted for both isospins and Cartesian 3-momenta, with $k_z$ and $k_{iz}$ given in Eq. \eqref{eq:kzdef}. The amplitude 
 $ \langle \sigma,\{\sigma_\ell \}; 
\tau,\{\tau_\ell\}; {\bf k},
\{{\bf k}_i\}|A;J_A,\pi_A,{\cal M}, T^A_z
\rangle$
is the wave function of an $A$-nucleon bound system, in momentum space, that is evaluated in the {\em instant form} (IF) of the Dirac relativistic Hamiltonian dynamics (see Ref. \cite{DelDotto:2016vkh} for A=3). One can demonstrate that the amplitude we need within the LF framework is related to the IF one by the proper number of Melosh rotations \cite{Melosh:1974cu} that transform  canonical spins into  LF ones. Due to the sum over the nucleon spins in Eq. \eqref{eq:momdis2},  the Melosh rotations reduce to unity, and we remain with canonical spins. This is a very important result, since the canonical spins  allow the use of the standard 
Clebsch-Gordan machinery.

The key point in our approach is: which is the wave function  of an $A$-nucleon bound system one has to use in Eq. \eqref{eq:momdis2} ? It is fundamental for our scope that it is obtained in a framework that fulfills the constraints imposed on the interacting mass operator by the BT construction of interacting generators of the Poincar\'e group. The necessary requirement is the dependence of the interactions on only scalar products among  intrinsic 3-momenta at disposal. One should notice that this is the case for Hamiltonians routinely adopted in a non relativistic framework, once we reinterpret the Cartesian 3-momenta there present, as the ones we have formally introduced through Eqs. \eqref{eq:general} and \eqref{eq:kzdef}. Moreover, as it is well-known for the two-nucleon case, which is relevant for  extracting information on the two-nucleon interaction from the phase-shifts (see, e.g., Ref. \cite{Lev:1993pfz}), one can formally embed  the {\em non relativistic} Schr\"odinger equation and the Lippmann-Schwinger one in a Poincar\'e covariant framework, once the BT construction is exploited.
In conclusion, within the  BT construction, one has for the interacting mass operator
\be
M[1,2,3,\dots,A]=M_0[1,2,3,\dots,A]+{\cal V}(|{\bf k}|^2; {\bf k}\cdot {\bf k}_j;{\bf k}_i\cdot {\bf k}_j)~,
\ee
or equivalently
\be
M^2[1,2,3,\dots,A]=M^2_0[1,2,3,\dots,A]+{\cal U}(|{\bf k}|^2; {\bf k}\cdot {\bf k}_j;{\bf k}_i\cdot {\bf k}_j).
\ee
For the two-body case one adopts the second expression. What we can do in practice is to use the  {\em non relativistic} wave function, expressed in terms of Jacobi momenta  of the $A$-nucleon bound system (for the three-body case see \cite{Pace:2022qoj,Pace:2020ned}), and obtain a Poincar\'e covariant description.  
{Lets us emphasize that this is the merit of} the BT construction and of the fit to the phase-shifts, which clearly incorporate the whole (relativistic) nature of the NN scattering.

Since we are considering DIS processes, the issue of macroscopic locality has to be taken into account, i.e., if the sub-clusters which compose a system are brought to infinity, the Poincar\'e generators of the system must reduce to  the sum of the sub-cluster Poincar\'e generators. It is worth noting that the packing operators which cooperate  for implementing   the macroscopic locality \cite{KP,sokolov1977relativistic}, are not considered in the present approximation for the description of the bound state. However, when we build up the LF spectral function we implement macroscopic locality through the tensor product of a plane wave for the knocked-out constituent  times the {fully interacting} intrinsic state for the $(A-1)$ spectator system, as allowed by the choice of LF  nonsymmetric intrinsic momenta (see Ref. \cite{DelDotto:2016vkh}). 

\section{Results}
The main ingredient for calculating the EMC ratio $R^A_{EMC}(x)$ is 
the light-cone momentum distributions for proton and neutron. In the actual application of our general formalism to  \hef, we used the $\alpha$-particle wave functions, obtained from i) the NN charge-dependent AV18+3N  \cite{Wiringa:1994wb,Pudliner:1995wk}  and ii) two versions of the Norfolk $\chi$EFT interactions, NVIa+3N and NVIb+3N, that generate three-nucleon forces within the same framework \cite{Piarulli:2016vel,Piarulli:2017dwd,Piarulli:2022ulk}.
It should be recalled that both NVIa and b interactions fit  the  NN phase shifts up to 125 MeV,  with $\chi^2 \sim 1$ (see Refs. \cite{Piarulli:2016vel,Piarulli:2017dwd,Piarulli:2022ulk}), and  the calculated \hef~binding energy is  $-28.298$  and $ -28.247$ MeV for NVIa+3N and b, respectively.
The short-range component of the 3N interactions  has two adjusted parameters, $C_E$ and  $c_D$. The first parameter, $c_E$, is fixed by reproducing   the \het~ and \tri~ binding energies, then the  \hef~ energy follows; the second one is fixed  by reproducing the matrix element of the   Gamow-Teller transition of \tri. Let us recall that the \hef~ wave functions corresponding to NVIa+3N and b interactions have been presented in Refs. 
\cite{Marcucci:2019hml} and  \cite{Piarulli:2016vel}, respectively. 

In addition to the light-cone distribution of \hef, we have  evaluated the ones for: i) the deuteron and  ii) \het~and \tri~ (the result of \het~ corresponding to AV18+3N is given in Ref. \cite{Pace:2022qoj}). The calculations,  obtained 
by using Eqs. \eqref{eq:f1xi} and \eqref{eq:momdis2}, are shown in Fig. \ref{fig:f1xi}:  in the left panel one finds   \hef~ and ${\rm ^2H}$ results ($Z=N$), while in the right panel the \het~ case ($Z\ne N$).  Notice that 
{the support, $0\le\xi\le 1$, is naturally obtained within our approach, and} both normalization and  momentum sum-rule, that amounts to $\langle \xi \rangle =1/A$,  have been numerically checked. The height of the peaks are almost the same, and their position is  at $\xi=1/A$. 
The tail of  \hef,  \het~and deuteron light-cone distributions corresponding to the AV18+3N interaction is bigger than the ones obtained from  the $\chi$EFT potentials. {Indeed, this feature is generated by the different content of {SRCs}, which affect the tail of the nucleon momentum distribution for the respective nuclei (see, e.g.,  \texttt{https://www.phy.anl.gov/theory/research/momenta/ }
for detailed calculations of both single and two-nucleon momentum distributions for light nuclei), and in turn the tail of the light-cone distributions, since low or high values of $\xi$ imply high values of $|k_z|$ \cite{LEV}.} 
Moreover, the  \hef~ tail for NVIb+3N is greater than the one for NVIa+3N, and the same ordering can be found also for the A=3 nuclei (in the right panel of Fig. \ref{fig:f1xi}, only the results for AV18+3N and NVIb+3N are shown, for the sake of readability), as well as for the deuteron.  Finally,  
{for increasing $A$ the $f_1(\xi)$ widths decrease, while the asymmetry with respect to the position of the peak increases.} {This can be understood by recalling that $k^+$ in Eq. \eqref{eq:kzdef} increases with $A$, for fixed values of $\xi$ (cf. Eq. \eqref{eq:general})}. {Hence, for $\xi>1/A$ one has an increasing $k_z$ for a growing $A$, and more and more large values of $|{\bf k}|$ come into play making faster the $f_1(\xi)$  decreasing. Differently, for {$\xi< 1/A $ the variable $k_z$ becomes negative and}
decreases in modulo for growing $A$, resulting in a $f_1(\xi)$ fall-off different from the one seen for $\xi>1/A$. {Interestingly, the asymmetry   of $f_1(\xi)$  has an impact on  the description  of the EMC effect for the studied light nuclei, and one could surmise that the same happens for heavier nuclei (cf. the extrema of integration in Eq. \eqref{eq:F2A}).  

In the following study of the EMC effect, it will be  interesting to look  for a possible signature of the  differences induced by the {SRCs}, as shown in the tails of the light-cone distributions.

\begin{figure}[htb]
\begin{center}
\includegraphics[width=6.7cm]{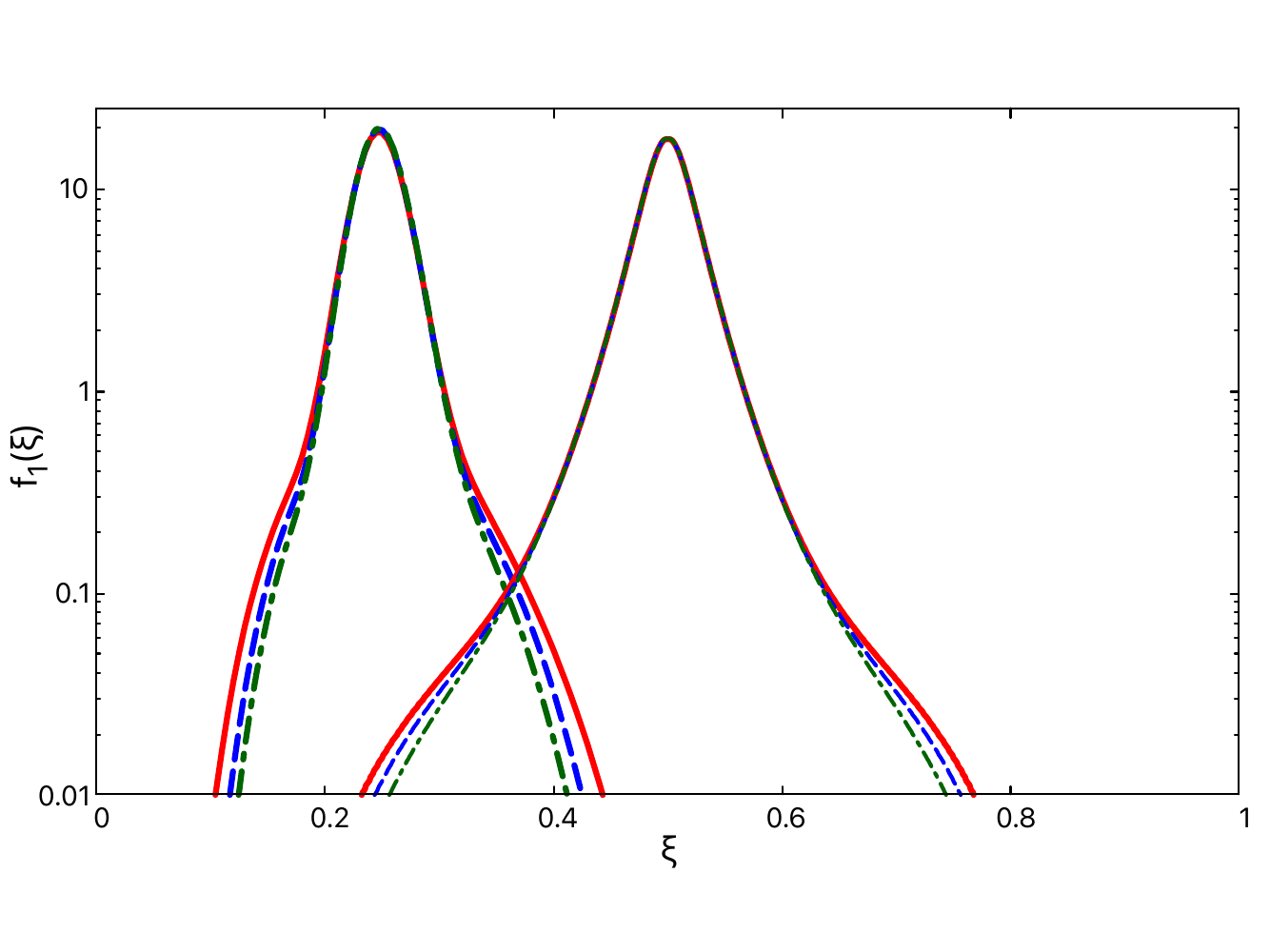}
\hspace{-0.2cm}\includegraphics[width=6.7cm]{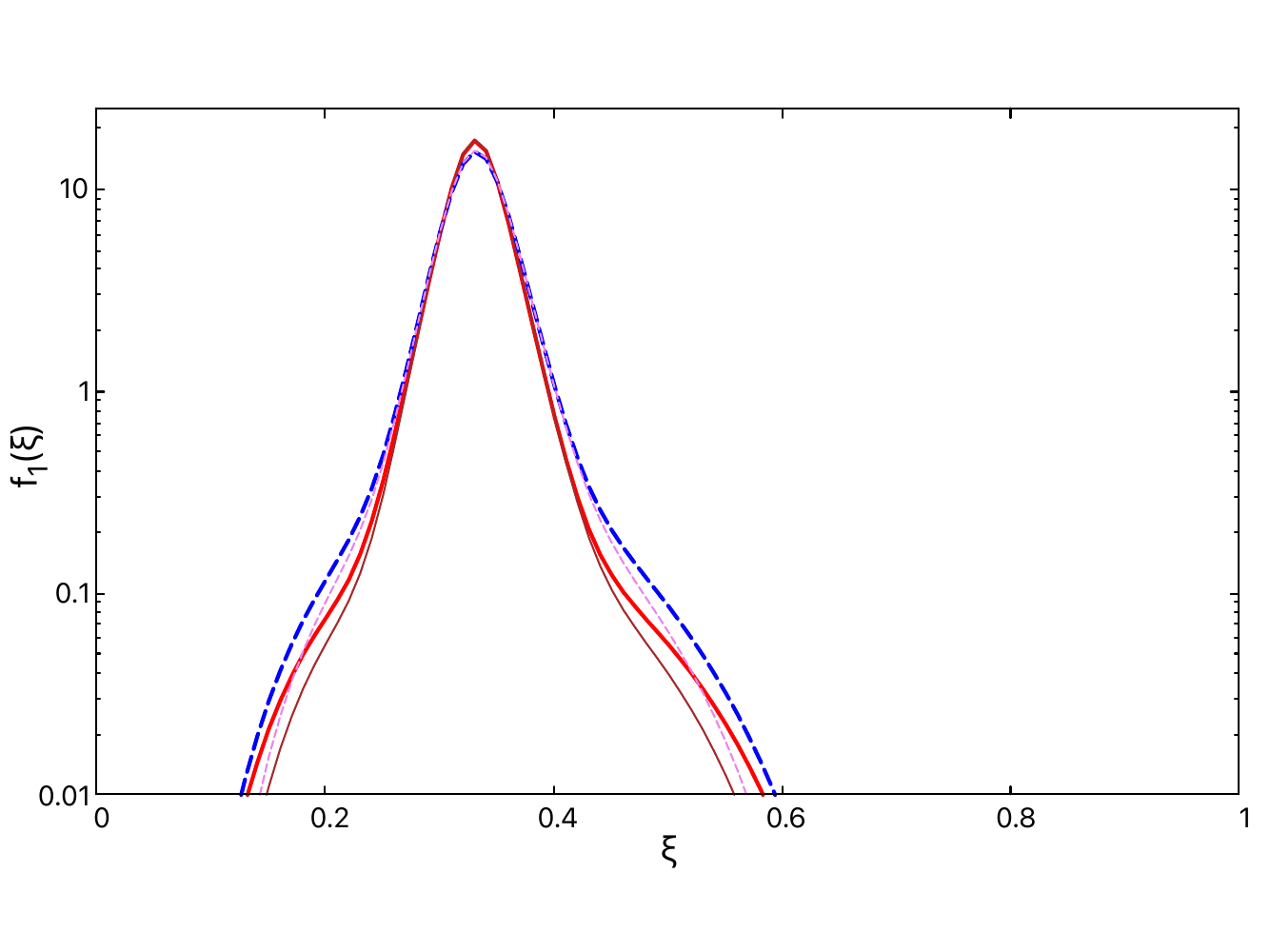}
\caption{ (Color online). Light-cone momentum distributions for \hef~,  \het~and ${\rm ^2H}$ obtained by using Eqs. \eqref{eq:f1xi} and \eqref{eq:momdis2}. Left panel. Solid lines:  \hef, peaked at $\xi=1/4$ and deuteron peaked at $\xi=1/2$. 
 Calculations have been obtained  by using   wave functions   corresponding to the AV18+3N  potential \cite{Wiringa:1994wb,Pudliner:1995wk}. Dashed lines: the same as the solid line, but 
corresponding to the $\chi$EFT interaction NVIb+3N \cite{Piarulli:2016vel,Piarulli:2017dwd,Piarulli:2022ulk}.
Dot-dashed lines: the same as the dashed lines, but  corresponding to the $\chi$EFT interaction NVIa+3N. Right panel. The same as the left panel but for  proton (solid lines) and neutron (dashed lines) in \het, peaked at $\xi=1/3$. Thick lines:  results obtained from AV18+3N in Ref. \cite{Pace:2022qoj}. Thin lines: results obtained from the $\chi$EFT interaction NVIb+3N \cite{Piarulli:2016vel,Piarulli:2017dwd,Piarulli:2022ulk}. 
}
\label{fig:f1xi}
\end{center}
\end{figure}

The EMC ratio has been calculated in the Bjorken limit and compared to the experimental data. Although the latter ones have finite values of the momentum transfer, one can safely adopt the Bjorken limit given the tiny dependence upon $Q^2$ \cite{Geesaman:1995yd,Cloet:2019mql,Arrington:2021vuu}.
Once $F^A_2(x)$, Eq. \eqref{eq:F2A}, is evaluated for \hef, one obtains the following EMC ratio 
\be
R^{^4He}_{EMC}(x)={R^{^4He}_2(x)\over R^D_2(x)}  ~,
\ee
where
\be
R^A_2(x)={F^A_2(x)\over Z~F^p_2(x)+(A-Z)~F^n_2(x)}~.
\ee
The results of  \hef~ and, for the sake of comparison, of \het~are shown in the left and the right panels of Fig. \ref{fig:EMCratio}, respectively. For both nuclei, it has been used the proton structure function $F^p_2(x)$ obtained by the SMC Coll. \cite{SpinMuonSMC:1997voo}, while $F^n_2(x)=r(x)~F^p_2(x)$, with $r(x)$  extracted from the MARATHON Coll. data \cite{MARATHON:2021vqu} in Ref. \cite{Pace:2022qoj}. It has to be pointed out that almost overlapping lines are obtained by using the ratio $r(x)$ provided by Ref. \cite{Valenty:2022gqm}, as well as $F^n_2(x)$ from the SMC Coll. \cite{SpinMuonSMC:1997voo}. This result strongly suggests that the  dependence on the ratio $F^n_2(x)/F^p_2(x)$ is largely under control. The lines shown in Fig. \ref{fig:EMCratio} have been calculated by exploiting three sets of \hef~ and \het~ wave functions, corresponding to i) the AV18+3N \cite{Wiringa:1994wb,Pudliner:1995wk} and ii) the $\chi$EFT interactions NVIa+3N {and NVIb+3N} \cite{Piarulli:2016vel,Piarulli:2017dwd,Piarulli:2022ulk}. 
It is clear that a formally correct use of the LF Hamiltonian dynamics plus BT construction and the reliable description of nuclei lead to predict  a ratio less than 1 in the range $0.2<x<0.75$, still taking into account only the dofs of the standard nuclear physics.  This result appears quite encouraging for achieving a quantitative separation between  {hadronic} dofs and partonic ones, and helps to progress in the path toward  understanding the transition from one  realm to another.
It is interesting to notice that the AV18+3N interaction, which gives rise to a higher momentum content and then more SRCs, produces a deeper minimum in the EMC ratio both for \hef~ and \het~ with respect to the two $\chi$EFT interactions which give almost the same results.
However, from Fig. \ref{fig:EMCratio}, one can see that 
the differences between the calculations with these  potentials 
are definitely less than the difference between data and theoretical  predictions around the dip, i.e. $x\sim  0.7$, leaving space for other effects. 

{One can notice} that the dip displacement of our calculations is not greatly affected by either the nuclear interaction or the ratio $F^n_2(x)/F^p_2(x)$, but it should be  driven by the treatment of the off-shellness of the interacting nucleon stricken by the virtual photon. } 
 \begin{figure}[htb]
\begin{center}
\includegraphics[width=6.8 cm]{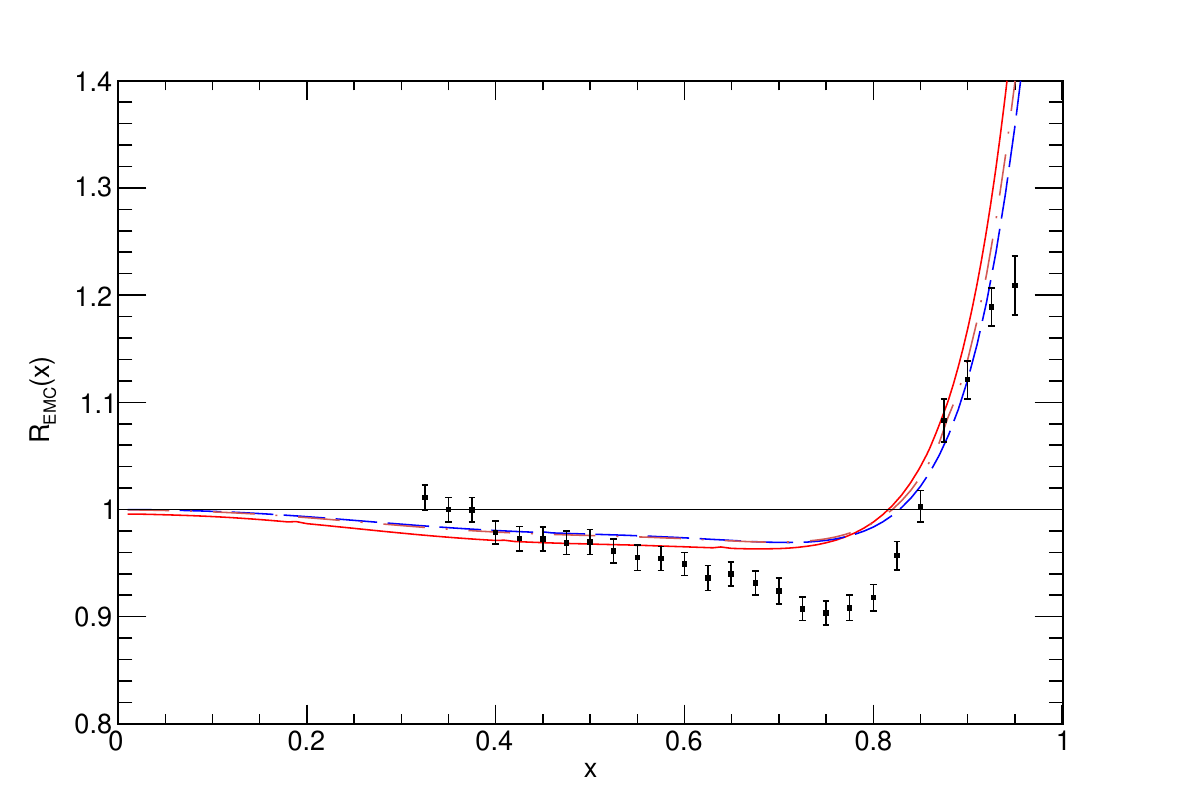}
\hspace{-0.4cm}\includegraphics[width=6.8 cm]{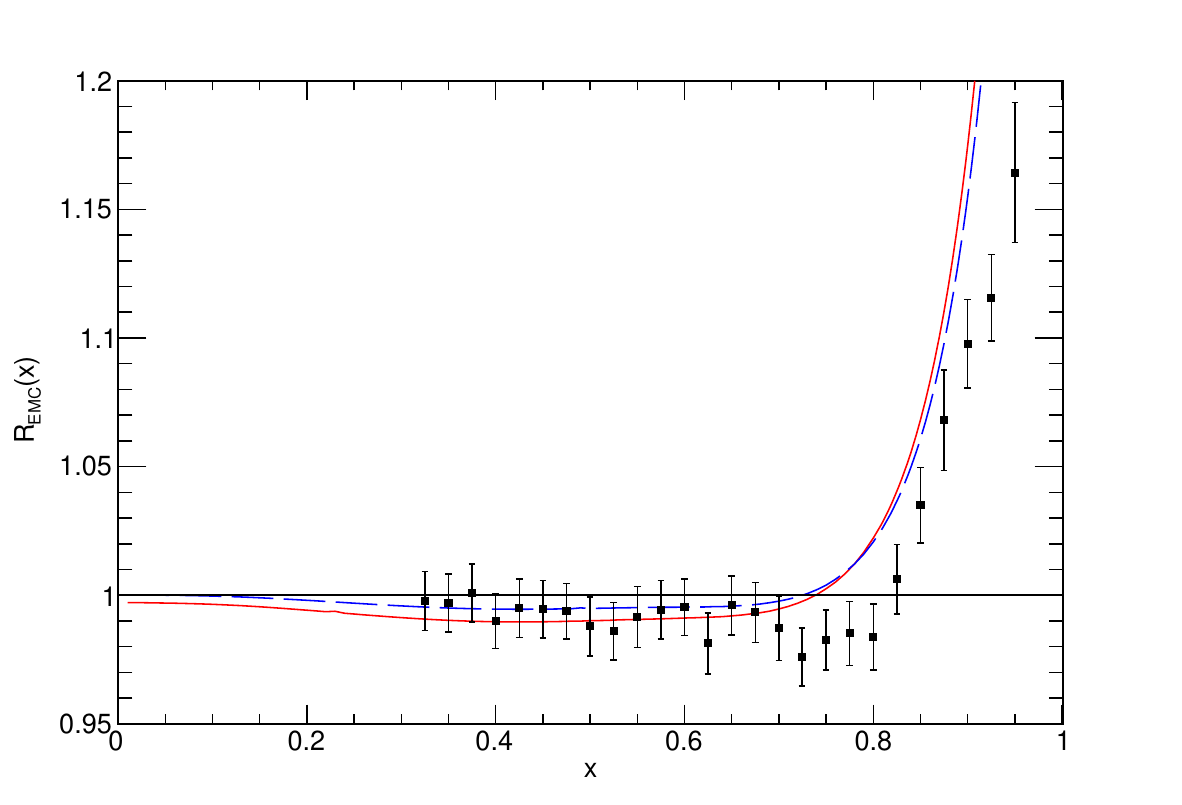}
\caption{Left panel. The EMC ratio for \hef. Solid line: our calculation with i) $F_2^p(x)$ of Ref. \cite{SpinMuonSMC:1997voo}, ii) $F_2^n(x) = r(x)~F_2^p(x)$, where $r(x)$ has been determined in Ref. \cite{Pace:2022qoj} (see Eq. (12)) by using the experimental data of the MARATHON Collaboration \cite{MARATHON:2021vqu} and iii)
the light-cone distribution corresponding to the wave function calculated in Ref. \cite{Marcucci:2019hml} with AV18+3N \cite{Wiringa:1994wb,Pudliner:1995wk}.  {Dashed line: the same as the solid one, but using    the  NVIa+3N interaction \cite{Piarulli:2016vel,Piarulli:2017dwd,Piarulli:2022ulk}. Dash-dotted line: the same as the dashed line but for  NVIb+3N.} N.B. The results obtained by using the {two} $\chi$EFT interactions are hardly distinguishable. Black squares: Jlab data of Ref. \cite{Seely:2009gt}. Right panel. {The EMC ratio for \het, using the wave functions from AV18+3N and NVIa+3N potentials (the results from NVIa+3N and NVIb+3N fully overlap)}. Black squares:  Jlab data of Ref. \cite{Seely:2009gt}, as reanalyzed in Ref. \cite{Kulagin:2010gd}.}
\label{fig:EMCratio}
\end{center}
\end{figure}

In view of future experimental data and for the sake of completeness, the EMC ratio for \tri~ is shown in Fig. \ref{fig:EMC_H3} for all the nuclear potentials we have discussed.
It is noteworthy that even for \tri~ the dependence of our uncertainties on the ratio $r(x)$ is very weak, as confirmed by the  result obtained by using  $r(x)$ given in   Ref. \cite{Valenty:2022gqm}, that overlaps the one from $r(x)$ of Ref. \cite{Pace:2022qoj}, shown in Fig. \ref{fig:EMC_H3}.
\begin{figure}[htb]
\begin{center}
\includegraphics[width=6.8 cm]{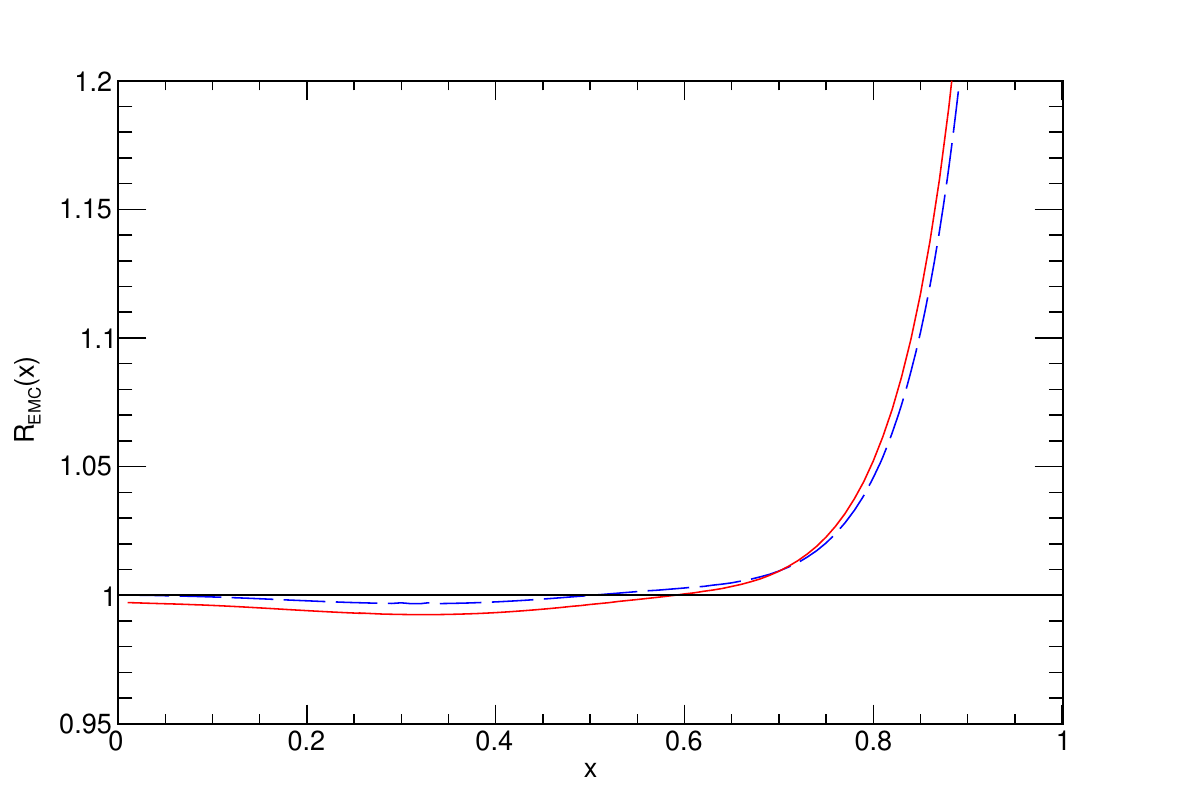}
\hspace{-0.3cm}\includegraphics[width=6.8 cm]{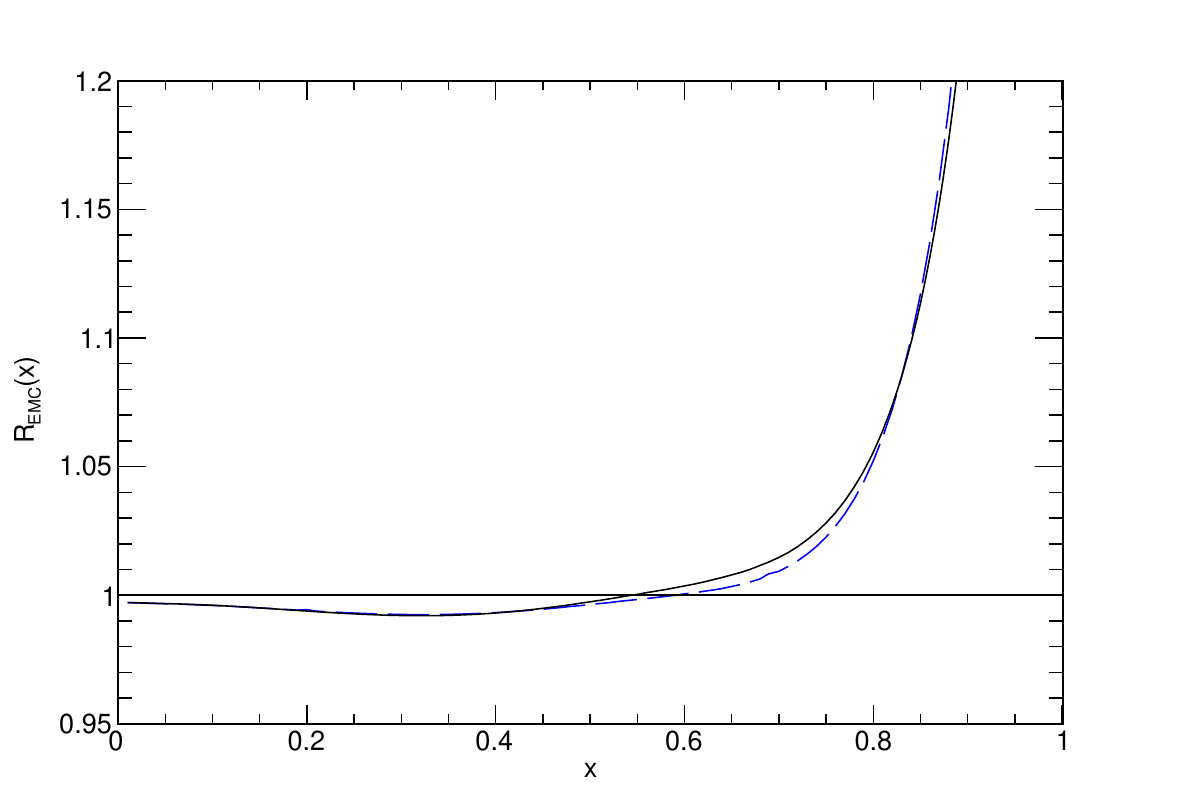}
\caption{The EMC ratio for  \tri.  Left panel.  
Solid line: our calculation with i) $F_2^p(x)$ of Ref. \cite{SpinMuonSMC:1997voo}, ii) $F_2^n(x) = r(x)~F_2^p(x)$, where $r(x)$ has been determined in Ref. \cite{Pace:2022qoj} (see Eq. (12)) by using the experimental data of the MARATHON Collaboration \cite{MARATHON:2021vqu} and iii)
the light-cone distribution corresponding to the wave function calculated in Ref. \cite{Marcucci:2019hml}  with AV18+3N \cite{Wiringa:1994wb,Pudliner:1995wk}. Dashed line: the same as the solid one, but with the outcome from NVIa+3N.   Right panel. Comparison with {two} different determination of $F^n_2(x)$, using the \tri~wave function from AV18+3N. Solid line: $F^n_2(x)$ from the  SMC Coll. \cite{SpinMuonSMC:1997voo}. Dashed line: $F^n_2(x)$ obtained through $r(x)$ of Ref. \cite{Pace:2022qoj}. }
\label{fig:EMC_H3}
\end{center}
\end{figure}

In Fig. \ref{fig:He4_CJ5TMC.pdf}, the impact of the proton structure function on the EMC ratio is analyzed by using two choices: i)  $F_2^p(x)$ of Ref. \cite{SpinMuonSMC:1997voo} and  ii) $F_2^p(x)$
of  Ref.  \cite{Accardi:2016qay},  the one which includes target mass corrections, keeping the same ratio $r(x)$ of Ref. \cite{Pace:2022qoj}. This study puts in evidence  the lack of knowledge  on the proton structure function for $x > 0.7$ (see  Sect. 18 in the  recent Particle Data Group \cite{Workman:2022ynf}), that however does not affect the typical range of $x$ where the EMC effect manifests.
\begin{figure}[htb]
\begin{center}
\includegraphics[width=6.8cm]{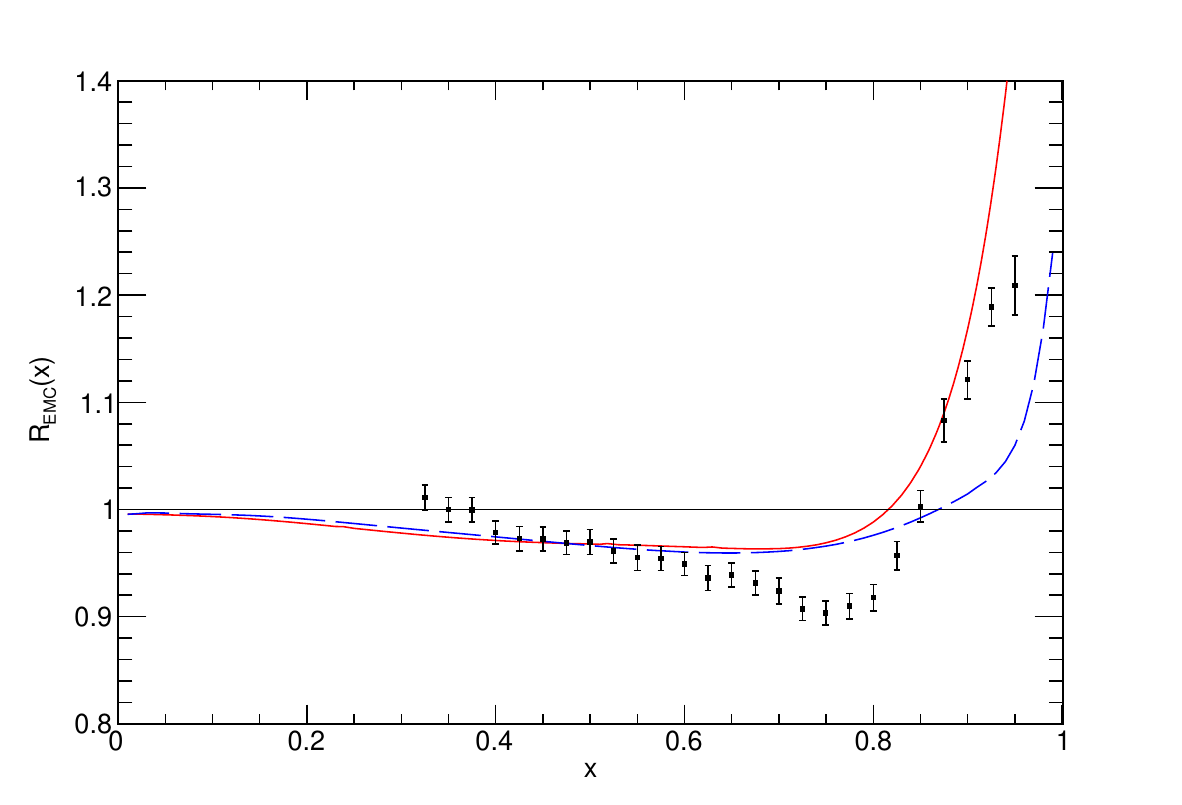}
\caption{EMC ratio in \hef~evaluated with different  proton structure functions and $F_2^n(x) = r(x)~F_2^p(x)$, where $r(x)$ has been determined in Ref. \cite{Pace:2022qoj}. Solid line:  $F_2^p(x)$ from  the SMC Coll. \cite{SpinMuonSMC:1997voo}. Dashed line:  $F_2^p(x)$ from  Ref.  \cite{Accardi:2016qay},  the one which includes target mass corrections.}
\label{fig:He4_CJ5TMC.pdf}
\end{center}
\end{figure}

\section{Conclusions}
We have extended our Poincar\'e covariant approach, based on the LF Hamiltonian dynamics supplemented by the Bakamjian-Thomas construction of the Poincar\'e generators (needed for determining a suitable mass operator for $A$ interacting nucleons), to the calculation of the EMC effect for any nuclei. 
Actual calculations have been performed for the first time for  \hef~ and \tri,  by using  wave  functions obtained from  modern nuclear interactions: i) Argonne V18 plus the 3N Urbana IX  and ii) two $\chi$EFT interactions: NVIa+3N and NVIb+3N. Moreover, also the \het~ has been re-analyzed by using wave functions corresponding to the $\chi$EFT interactions.

This is a substantial step forward for providing a reliable baseline, which 
retains only the dofs of the  standard nuclear physics, but within a Poincar\'e covariant framework. For the very light nuclei, the impact on $R^A_{EMC}(x)$ of the different short-range correlation content, generated by  the  realistic nuclear potentials we have considered,  has been quantitatively studied, and it results quite smaller than the EMC effect by itself. 
Moreover, the uncertainty on the ratio between the neutron and proton structure functions appears to be negligible, while the lack of knowledge on $F^p_2(x)$ for $x>0.7$  clearly calls for new experimental campaigns.

In principle, our analysis could help in ascribing the  deviation of the EMC ratio from the proposed baseline,  to exotic phenomena involving necessarily partonic dofs, not included in a standard nuclear description, and possibly  initiates a systematic and quantitative analysis of the transition from 
hadronic to QCD dofs in  light nuclei.

\bibliography{EMC_A_arxiv.bib}
\end{document}